\documentstyle[aps,epsf,floats]{revtex}            % GALLEY STYLE
\def\bq{\begin{equation}}
\def\eq{\end{equation}}
\def\ba{\begin{eqnarray}}
\def\ea{\end{eqnarray}}

\begin{document}
\thispagestyle{empty}

\newcommand{\sla}[1]{/\!\!\!#1}

\renewcommand{\small}{\normalsize} %% Remove for Phys.Rev.

\preprint{
\font\fortssbx=cmssbx10 scaled \magstep2
\hbox to \hsize{
\hskip.5in \raise.1in\hbox{\fortssbx University of Wisconsin - Madison}
\hfill\vtop{\hbox{\bf MADPH-99-1112}
            \hbox{May 1999}} }
}

\title{\vspace{.5in}
Observing $H \rightarrow W^{(*)} W^{(*)} \to e^{\pm} \mu^{\mp} \sla{p_T}$ 
\\ in weak boson fusion with dual forward jet tagging 
\\ at the CERN LHC
}
\author{D.~Rainwater \ and D.~Zeppenfeld\\[3mm]}
\address{
Department of Physics, University of Wisconsin, Madison, WI 53706
}
\maketitle
\begin{abstract}
Weak boson fusion promises to be a copious source of intermediate mass 
Standard Model Higgs bosons at the LHC. The additional very energetic 
forward jets in these events provide for powerful background suppression 
tools. We analyze the $H \to W^{(*)}W^{(*)} \to e^{\pm} \mu^{\mp} \sla{p_T}$ 
decay mode for a Higgs boson mass in the 130-200~GeV range. A parton level 
analysis of the dominant backgrounds (production of $W$ pairs, $t\bar{t}$ 
and $Z \to \tau\tau$ in association with jets) demonstrates that this 
channel allows the observation of $H \to W^{(*)}W^{(*)}$ in a virtually 
background-free environment, yielding a significant Higgs boson signal with 
an integrated luminosity of 5~fb$^{-1}$ or less. Weak boson fusion achieves 
a much better signal to background ratio than inclusive 
$H\to e^{\pm} \mu^{\mp} \sla{p_T}$ and is therefore the most promising 
search channel in the 130-200~GeV mass range.
\end{abstract}

%%%%%%%%%%%%%%%%%%%%%%%%%%%%%%%%%%%%%%%%%%%%%%%%%%%%%%%%%%%%%%%%%%%%%%%%%%%%%%%%
%%%%%%%%%%%%%%%%%%%%%%%%%%%%%%%%%%%%%%%%%%%%%%%%%%%%%%%%%%%%%%%%%%%%%%%%%%%%%%%%

\section{Introduction}
\label{sec:intro}

The search for the Higgs boson and, hence, for the origin of electroweak 
symmetry breaking and fermion mass generation, remains one of the premier 
tasks of present and future high energy physics experiments. Fits to 
precision electroweak (EW) data have for some time suggested a relatively 
small Higgs boson mass, of order 100~GeV~\cite{EWfits}. This is one of the 
reasons why the search for an intermediate mass Higgs boson is particularly 
important~\cite{reviews}. 

For the intermediate mass range, most of the literature has focussed on 
Higgs boson production via gluon fusion~\cite{reviews} and 
$t\bar{t}H$~\cite{ttH} or $WH(ZH)$~\cite{WH} associated production. Cross 
sections for Standard Model (SM) Higgs boson production at the LHC are 
well-known~\cite{reviews}, and while production via gluon fusion has the 
largest cross section by almost one order of magnitude, there are substantial 
QCD backgrounds. A search for the very clean four-lepton signature from 
$H\to ZZ$ decay can find a Higgs boson in the mass region 
$m_H \gtrsim 130$~GeV, but due to the small branching fraction of this mode 
very large integrated luminosities, up to 100~fb~$^{-1}$ or more, are 
required. One can search for $gg\to H$ via 
$H\to W^{(*)}W^{(*)} \to e^{\pm} \mu^{\mp} \sla{p_T}$ decays with much lower 
luminosity~\cite{glo_will_hww,DittDrein,bpz_minijet}, but with lower 
signal-to-background ratios.

The second largest production cross section is predicted for weak-boson 
fusion (WBF), $qq \to qqVV \to qqH$. These events contain additional 
information in their observable quark jets. Techniques like forward jet 
tagging~\cite{Cahn,BCHP,DGOV} can then be exploited to significantly reduce 
the backgrounds. WBF and gluon fusion nicely complement each other: 
together they allow for a measurement of the $t\bar{t}H/WWH$ coupling ratio.

Another feature of the WBF signal is the lack of color exchange between the 
initial-state quarks. Color coherence between initial- and final-state gluon 
bremsstrahlung leads to suppressed hadron production in the central region, 
between the two tagging-jet candidates of the signal~\cite{bjgap}. This is in 
contrast to most background processes, which typically involve color exchange 
in the $t$-channel and thus lead to enhanced hadronic activity between the 
tagging jets. We exploit these features, via a veto of soft jet activity 
in the central region~\cite{bpz_minijet}.

While some attention has been given to intermediate-mass $H\to W^{(*)}W^{(*)}$ 
searches at the LHC in the framework of gluon 
fusion~\cite{glo_will_hww,DittDrein}, 
production via weak boson fusion for the same decay mode has not yet been 
discussed in the literature. 
Thus, we provide a first analysis of intermediate-mass 
$VV\to H\to W^{(*)}W^{(*)}$ at the LHC (and of the main physics and 
reducible backgrounds) which demonstrates the feasibility of Higgs boson 
detection in this channel, with very low luminosity. 
$H\to W^{(*)}W^{(*)}$ event characteristics are analyzed for dual leptonic 
decays to $e^{\pm}\mu^{\mp}$ only, to avoid backgrounds from 
$Z,\gamma \to e^+e^-,\mu^+\mu^-$.

Our analysis is a parton-level Monte Carlo study, using full tree-level 
matrix elements for the WBF Higgs signal and the various backgrounds. 
In Section~\ref{sec:calc} we describe our calculational tools, the 
methods employed in the simulation of the various processes, and important 
parameters. Extra minijet activity is simulated by adding the emission of one 
extra parton to the basic signal and background processes. Generically we call 
the basic signal process (with its two forward tagging jets) and the 
corresponding background calculations ``2-jet'' processes, and refer to the 
simulations with one extra parton as ``3-jet'' processes. In
Section~\ref{sec:analysis}, using the 2-jet programs for the backgrounds, we 
demonstrate forward jet tagging, a $b$ veto and other important cuts which 
combine to yield an $\approx \;$2/1 to 1/2 signal-to-background (S/B) ratio, 
depending on the Higgs mass. 

In Section~\ref{sec:minijet} we analyze the different minijet patterns in signal 
and background, using both the truncated shower approximation (TSA)~\cite{TSA} 
to regulate the cross sections, and the gluon exponentiation model to estimate 
the minijet multiplicity~\cite{CDFjets}. By exploiting the two most important 
characteristics of the extra radiation, its angular distribution and its 
hardness, the QCD backgrounds can be suppressed substantially by a veto on 
extra central jet emission. Within the TSA and exponentiation models, 
probabilities are estimated for vetoing signal and background events, and are 
combined with the production cross sections of the previous section to predict 
signal and background rates in Table~\ref{summary}. 
These rates demonstrate the feasibility of extracting a very low background 
$H\to W^{(*)}W^{(*)}$ signal at the LHC.

Our signal selection is not necessarily optimized yet. The variables we 
identify for cuts are the most distinctive, but deserve a multivariate 
analysis with detector simulation. We do construct an additional variable in 
Section~\ref{sec:disc} which is not used for cuts, but rather can be used to 
extract the Higgs boson mass from the final event sample.

%%%%%%%%%%%%%%%%%%%%%%%%%%%%%%%%%%%%%%%%%%%%%%%%%%%%%%%%%%%%%%%%%%%%%%%%%%%%%%%%
%%%%%%%%%%%%%%%%%%%%%%%%%%%%%%%%%%%%%%%%%%%%%%%%%%%%%%%%%%%%%%%%%%%%%%%%%%%%%%%%

\section{Calculational Tools}
\label{sec:calc}

We simulate $pp$ collisions at the CERN LHC, $\protect\sqrt{s} = 14$~TeV. 
All signal and background cross sections are determined in terms of full 
tree level matrix elements for the contributing subprocesses and are 
discussed in more detail below.

For all our numerical results we have chosen ${\rm sin}^2\theta_W = 0.2315$,
$M_Z = 91.19$~GeV, and $G_F = 1.16639\cdot 10^{-5}\;{\rm GeV}^{-2}$, which
translates into $M_W = 79.94$~GeV and $\alpha(M_Z) = 128.74$ when using
the tree-level relations between these input parameters. This value of $M_W$
is somewhat lower than the current world average of $\approx 80.35$ GeV.
However, this difference has negligible effects on all cross sections, e.g.
the $qq\to qqH$ signal cross section varies by about $0.5\%$ between these two
$W$ mass values. The tree level relations between the input parameters are
kept in order to guarantee electroweak gauge invariance of all amplitudes.
For all QCD effects, the running of the strong coupling constant is
evaluated at one-loop order, with $\alpha_s(M_Z) = 0.118$. We employ
CTEQ4L parton distribution functions~\cite{CTEQ4_pdf} throughout. Unless
otherwise noted the factorization scale is chosen as $\mu_f =$ min($p_T$)
of the defined jets.

%%%%%%%%%%%%%%%%%%%%%%%%%%%%%%%%%%%%%%%%%%%%%%%%%%%%%%%%%%%%%%%%%%%%%%%%%%%%%%%%

\subsection{The $qq\to qqH(g)$ signal process}
\label{sec:qqH}

The signal can be described, at lowest order, by two single-Feynman-diagram
processes, $qq \to qq(WW,ZZ) \to qqH$, {\em i.e.} $WW$ and $ZZ$ fusion where 
the weak bosons are emitted from the incoming quarks~\cite{qqHorig}. Because of 
the small Higgs boson width in the mass range of interest, these events can 
reliably be simulated in the narrow width approximation. 
From previous studies of $H\to\gamma\gamma$~\cite{RZ_gamgam} and 
$H\to\tau\tau$~\cite{RZ_tautau} decays in weak boson fusion we know several 
features of the signal, which can be exploited here also: 
the centrally produced Higgs boson tends to yield central decay products 
(in this case $W^+W^-$), and the two quarks enter the detector at large 
rapidity compared to the $W$'s and with transverse momenta in the 
20 to 100 GeV range, thus leading to two observable forward tagging jets.

For the study of a central jet veto, we utilize the results of previous 
studies where we simulated the emission of at least one extra 
parton~\cite{RZ_tautau,RSZ_vnj}. This was achieved by calculating the 
cross sections for the process $qq\to qqHg$, {\em i.e.} weak boson fusion with 
radiation of an additional gluon, and all crossing related processes. 

An important additional tool for distinguishing the 
$H\to e^\pm\mu^\mp\sla{p_T}$ 
signal from various backgrounds is the anti-correlation of the $W$ spins, as 
pointed out in Ref.~\cite{DittDrein}. This is due to the preservation of 
angular momentum in the decay of the spin-0 Higgs boson. Of course, we can 
observe only the angular distributions of the charged decay leptons, but this 
is sufficient. The decay rate is proportional to 
$(p_{\ell^-}\cdot p_{\nu})(p_{\ell^+}\cdot p_{\bar\nu})$. In the rest frame of 
the Higgs boson, in which the $e^-\bar\nu$ or $e^+\nu$ pairs are emitted 
back-to-back for $W^+W^-$ production at threshold, this product is a maximum 
for the charged leptons being emitted parallel. This characteristic is 
preserved and even enhanced when boosted to the lab frame, as the Higgs 
boson in weak boson fusion is typically emitted with $p_T \approx 60-120$~GeV.

%%%%%%%%%%%%%%%%%%%%%%%%%%%%%%%%%%%%%%%%%%%%%%%%%%%%%%%%%%%%%%%%%%%%%%%%%%%%%%%%

\subsection{The QCD $t \bar{t} + jets$ backgrounds}
\label{sec:ttnj}

Given the H decay signature, the main physics background to our 
$e^{\pm}\mu^{\mp}\sla p_T$ signal arises from $t\bar{t} + jets$ production, 
due to the large top production cross section at the LHC and because 
the branching ratio $B(t\to Wb)$ is essentially $100\%$. 

The basic process we consider is $pp\to t\bar{t}$, which can be either $gg$- 
or $q\bar{q}$-initiated, with the former strongly dominating at the LHC. 
QCD corrections to this lead to additional real parton emission, {\em i.e.} 
to $t\bar{t} + j$ events. Relevant subprocesses are 
\bq\label{QCD_tt}
g q  \to t \bar{t} q \, , \qquad g \bar{q} \to t \bar{t} \bar{q} \, , \qquad 
q \bar{q} \to t \bar{t} g \, , \qquad g g \to t \bar{t} g \, ,
\eq
and the subprocesses for $t\bar{t} + jj$ events can be obtained similarly. 
For the case of no additional partons, the $b$'s from the decaying top quarks 
may be identified as the tagging jets. In this case, calculating the cross 
section for $t\bar{t} + j$ where the $b$'s are explicitly identified as the 
tagging jets serves to estimate the effect of additional soft parton emission, 
{\em i.e.} minijet activity in the central detector; this is described 
in detail in Sec.~\ref{sec:minijet}. At the same time, we can identify a 
distinctly different, perturbative region of phase space, where the 
final-state light quark or gluon gives rise to one tagging jet, and one of 
the two decay $b$'s is 
identified as the other tagging jet. In this case, $t\bar{t} + jj$ may be used 
to estimate minijet activity for the hard process $pp\to t\bar{t} + j$. 
Finally, there is a third distinct region of phase space, for the perturbative 
hard process $pp\to t\bar{t} + jj$, where the final state light quarks or 
gluons are the two tagging jets.

Thus, the ``$t\bar{t}j$'' and ``$t\bar{t}jj$'' calculations serve a dual 
purpose: to obtain the cross sections for the contribution of the perturbative 
processes where light quark or gluon jets lie in the region of phase space 
where they are experimentally identified as far-forward/backward tagging jets; 
and to estimate the additional QCD radiation patterns for the next-lower-order 
perturbative $t\bar{t} + jets$ process. The $t\bar t$ and $t\bar tj$ matrix 
elements were constructed using MadGraph~\cite{Madgraph}, while the 
$t\bar{t}jj$ matrix elements are from Ref.~\cite{Stange}.

Decays of the top quarks and $W$'s are included in the matrix elements; 
however, while the $W$'s are allowed to be off-shell, the top quarks are 
required to be on-shell. Energy loss from $b\to\ell\nu X$ is included to 
generate more accurate $\sla{p}_T$ distributions. In all cases, the 
factorization scale is chosen as $\mu_f =$ min($E_T$) of the massless 
partons/top quarks. The overall strong coupling constant factors are taken 
as $(\alpha_s)^n = \prod_{i=1}^n \alpha_s(E_{T_i})$, where the product runs 
over all light quarks, gluons and top quarks; {\em i.e.} the transverse 
momentum of each additional parton is taken as the relevant scale for its 
production, irrespective of the hardness of the underlying scattering event. 
This procedure guarantees that the same $\alpha_s^2$ factors are used for 
the hard part of a $t\bar{t} + jets$ event, independent of the number of 
additional minijets, and at the same time the 
small scales relevant for soft-parton emission are implemented.

%%%%%%%%%%%%%%%%%%%%%%%%%%%%%%%%%%%%%%%%%%%%%%%%%%%%%%%%%%%%%%%%%%%%%%%%%%%%%%%%

\subsection{The QCD $WW+jj$ background}

The next obvious background arises from real-emission QCD corrections 
to $W^+W^-$ production. For $W^+ W^- jj$ events these background processes 
include~\cite{VVjj}
\bq\label{QCD_WW}
q g \to q g W^+ W^- \, , \qquad  q q' \to q q' W^+ W^- \, ,
\eq
which are dominated by $t$-channel gluon exchange, and all crossing
related processes, such as
\bq
q \bar{q} \to g g W^+ W^- \, , \qquad g g \to q \bar{q} W^+ W^- \;.
\eq
We call these processes collectively the ``QCD $WWjj$'' background. We do not 
calculate cross sections for the corresponding $WW+3$-jet processes, but 
instead follow the results of our analysis of the radiation patterns of QCD 
$Z + jets$ processes, detailed in Sec.~\ref{sec:minijet}, and apply those 
results here to estimate minijet veto probabilities.

The factorization scale is chosen as for the Higgs boson signal. The strong 
coupling constant factor is taken as 
$(\alpha_s)^2 = \alpha_s(p_{T_1})\alpha_s(p_{T_2})$, {\em i.e.}, the 
transverse momentum of each additional parton is taken as the relevant scale 
for its production. Variation of the scales by a factor 2 or $1\over 2$ reveals 
scale uncertainties of $\approx 35\%$, however, which emphasizes the need for 
experimental input or NLO calculations.

The $WW$ background lacks the marked anti-correlation of $W$ spins seen in 
the signal. As a result the momenta of the charged decay leptons will be 
more widely separated than in $H\to W^{(*)}W^{(*)}$ events.

%%%%%%%%%%%%%%%%%%%%%%%%%%%%%%%%%%%%%%%%%%%%%%%%%%%%%%%%%%%%%%%%%%%%%%%%%%%%%%%%

\subsection{The EW $WW+jj$ background}

These backgrounds arise from $W^+W^-$ bremsstrahlung in
quark--(anti)quark scattering via $t$-channel electroweak boson exchange,
with subsequent decay $W^+W^-\to\ell^+\ell^-\sla p_T$:
\bq
qq' \to qq' W^+W^-
\label{eq:EW_WW}
\eq
Na\"{\i}vely, this EW background may be thought of as suppressed compared 
to the analogous QCD process in Eq.~(\ref{QCD_WW}). However, it includes 
electroweak boson fusion, $VV \to W^+W^-$ via $s$- or $t$-channel 
$\gamma/Z$-exchange or via $VVVV$ 4-point vertices, which has a momentum 
and color structure identical to the signal. Thus, it cannot easily be 
suppressed via cuts.

The matrix elements for these processes were constructed using 
MadGraph~\cite{Madgraph}. We include charged-current (CC) and neutral-current 
(NC) processes, but discard s-channel EW boson and t-channel quark exchange 
processes as their contribution was found to be $\approx 1\%$ only, while 
adding significantly to the CPU time needed for the calculation. In general, 
for the regions of phase space containing far-forward and -backward tagging 
jets, s-channel processes are severely suppressed. We refer collectively to 
these processes as the ``EW $WWjj$'' background. Both $W$'s are allowed to 
be off-shell, and all off-resonance graphs are included. In addition, the 
Higgs boson graphs must be included to make the calculation well-behaved at 
large $W$-pair invariant masses. However, these graphs include our signal 
processes and might lead to double counting. Thus, we set $m_H$ to 60~GeV in 
the EW $WWjj$ background to remove their contribution. A clean separation of 
the Higgs boson signal and the EW $WWjj$ background is possible because 
interference effects between the two are negligible for the Higgs boson mass 
range of interest.

Again we will need an estimate of additional gluon radiation patterns. 
This was first done for EW processes in Ref.~\cite{DZ_IZ_minijet}, but for  
different cuts on the hard process, and again for EW $\tau\tau jj$ processes in 
Ref.~\cite{RZ_tautau}. We reanalyze the EW $\tau\tau jj$ case in 
Sec.~\ref{sec:minijet} and directly apply the resulting minijet emission 
probabilities here. The EW $\tau\tau jj$ and EW $WWjj$ backgrounds are 
quite similar kinematically, which justifies the use of the same veto 
probabilities for central jets. 

%%%%%%%%%%%%%%%%%%%%%%%%%%%%%%%%%%%%%%%%%%%%%%%%%%%%%%%%%%%%%%%%%%%%%%%%%%%%%%%%

\subsection{The QCD and EW $\tau^+\tau^-$ backgrounds}
\label{sec:tau}

The leptonic decay of $\tau$'s provides a source of electrons, muons and 
neutrinos which can be misidentified as $W$ decays. Thus, we need to study 
real-emission QCD corrections to the Drell-Yan process 
$q\bar{q} \to (Z,\gamma) \to \tau^+\tau^-$. 
For $\tau^+ \tau^- jj$ events these background processes include~\cite{Kst}
\bq\label{QCD_tau}
q g \to q g \tau^+ \tau^- \, , \qquad  q q' \to q q' \tau^+ \tau^- \, ,
\eq
which are dominated by $t$-channel gluon exchange, and all crossing-related 
processes, such as
\bq
q \bar{q} \to g g \tau^+ \tau^- \, , \qquad g g \to q \bar{q} \tau^+ \tau^- \;.
\eq
All interference effects between virtual photon and $Z$-exchange are included.
We call these processes collectively the ``QCD $\tau\tau jj$'' background. The 
cross sections for the corresponding $\tau\tau+3$-jet processes, which we need 
for our modeling of minijet activity in the QCD $\tau\tau jj$ background, have 
been calculated in Refs.~\cite{HZ,BHOZ,BG}. Similar to the treatment of the 
signal processes, we use a parton-level Monte-Carlo program based on the work 
of Ref.~\cite{BHOZ} to model the QCD $\tau\tau jj$ and $\tau\tau jjj$ 
backgrounds.

From our study of $H\to\tau\tau$ in weak boson fusion~\cite{RZ_tautau}, we 
know that the EW (t-channel weak boson exchange) cross section will be 
comparable to the QCD cross section in the phase space region of interest. 
Thus, we consider those processes separately, in a similar manner as for the 
EW $WWjj$ contribution. We use the results of Ref.~\cite{CZ_gap} for modeling 
the EW $\tau\tau jj$ background.

The dual leptonic decays of the $\tau$'s are simulated by multiplying the 
$\tau^+\tau^-jj$ cross section by a branching ratio factor of $(0.3518)^2/2$ 
and by implementing collinear tau decays with helicity correlations included as 
in our previous analysis of $H\to\tau\tau$~\cite{RZ_tautau}.

%%%%%%%%%%%%%%%%%%%%%%%%%%%%%%%%%%%%%%%%%%%%%%%%%%%%%%%%%%%%%%%%%%%%%%%%%%%%%%%%

\subsection{Detector resolution}

The QCD processes discussed above lead to steeply falling jet transverse 
momentum distributions. As a result, finite detector resolution can have a 
sizable effect on cross sections. These resolution effects are taken into 
account via Gaussian smearing of the energies of jets/$b$'s and charged 
leptons.  We use 
\bq
{\triangle{E} \over E} =
{5.2 \over E} \oplus {0.4 \over {\sqrt E}} \oplus .009 \; ,
\eq
for jets (with individual terms added in quadrature), based on ATLAS 
expectations~\cite{CMS-ATLAS}. For charged leptons we 
use
\bq
{\triangle{E} \over E} = 2\% \; .
\eq

In addition, finite detector resolution leads to fake 
missing-transverse-momentum in events with hard jets. An ATLAS 
analysis~\cite{ATLAS_tau} showed 
that these effects are well parameterized by a Gaussian distribution of 
the components of the fake missing transverse momentum vector, 
$\vec\sla p_T$, with resolution 
\bq
\sigma(\sla p_x,\sla p_y) = 0.46 \cdot \sqrt{\sum{E_{T,had}}} \; ,
\eq
for each component. In our calculations, these fake missing transverse 
momentum vectors are added linearly to the neutrino momenta.

%%%%%%%%%%%%%%%%%%%%%%%%%%%%%%%%%%%%%%%%%%%%%%%%%%%%%%%%%%%%%%%%%%%%%%%%%%%%%%%%
%%%%%%%%%%%%%%%%%%%%%%%%%%%%%%%%%%%%%%%%%%%%%%%%%%%%%%%%%%%%%%%%%%%%%%%%%%%%%%%%

\section{Higgs signal and real $W^+W^-$ backgrounds}
\label{sec:analysis}

The $qq\to qqH, \; H\to W^{(*)}W^{(*)} \to e^\pm \mu^\mp \nu \bar{\nu}$ dual 
leptonic decay signal is characterized by two forward jets and the $W$ decay 
leptons ($e,\mu$). Before discussing background levels and further 
details like minijet radiation patterns, we need to identify the search region 
for these hard $Hjj$ events. The task is identical to the Higgs searches in 
$qq\to qqH,\;H\to\gamma\gamma,\tau\tau$ which were considered 
previously~\cite{RZ_gamgam,RZ_tautau}. We can thus adopt the strategy of these 
earlier analyses and start out by discussing a basic level of cuts on the 
$qq\to qqH,\;H\to W^{(*)}W^{(*)}$ signal. Throughout this section we assume a 
Higgs mass of $M_H = 160$~GeV, but we do not optimize cuts for this mass.

The minimum acceptance requirements ensure that the two jets and two charged 
leptons are observed inside the detector (within the hadronic and 
electromagnetic calorimeters, respectively), and are well-separated from each 
other:
\ba
\label{eq:basic}
& p_{T_j} \geq 20~{\rm GeV} \, ,\qquad |\eta_j| \leq 5.0 \, ,\qquad 
\triangle R_{jj} \geq 0.7 \, , \nonumber\\
& p_{T_\ell} \geq 20~{\rm GeV} \, ,\qquad
|\eta_{\ell}| \leq 2.5 \, , \qquad \triangle R_{j\ell} \geq 0.7 \, .
\ea
A feature of the QCD $WWjj$ background is the generally higher
rapidity of the $W$'s as compared to the Higgs signal: weak boson 
bremsstrahlung occurs at small angles with respect to the parent quarks,
producing $W$'s forward of the jets. Thus, we also require both $\ell$'s to 
lie between the jets with a separation in pseudorapidity 
$\triangle \eta_{j,\ell} > 0.7$, and the jets to occupy opposite hemispheres:
\bq
\label{eq:lepcen}
\eta_{j,min} + 0.7 < \eta_{\ell_{1,2}} < \eta_{j,max} - 0.7 \, , \qquad
\eta_{j_1} \cdot \eta_{j_2} < 0
\eq
Finally, to reach the starting point for our consideration of the signal and 
various backgrounds, a wide separation in pseudorapidity is required between 
the two forward tagging jets,
\bq
\label{eq:gap}
\triangle \eta_{tags} = |\eta_{j_1}-\eta_{j_2}| \geq 4.4 \, ,
\eq
leaving a gap of at least 3 units of pseudorapidity in which the $\ell$'s can 
be observed. This technique to separate weak boson scattering from various 
backgrounds is 
well-established~\cite{bpz_minijet,Cahn,BCHP,DGOV,RZ_gamgam,RZ_tautau,RSZ_vnj},
in particular for heavy Higgs boson searches. Line 1 of Table~\ref{WW_data} 
shows the effect of these cuts on the signal and backgrounds for a SM Higgs 
boson of mass $m_H = 160$~GeV. Overall, about $28\%$ of all 
$H\to W^{(*)}W^{(*)} \to e^\pm \mu^\mp \nu \bar{\nu}$ events generated in weak 
boson fusion are accepted by the cuts of Eqs.~(\ref{eq:basic}-\ref{eq:gap}) 
(for $m_H = 160$~GeV). 

Somewhat surprisingly, the EW $WWjj$ background rate reaches 2/3 of the QCD 
$WWjj$ background rate already at this level. This can be explained by the 
contribution from $W,Z,\gamma$ exchange and fusion 
processes which can produce central $W$ pairs and are therefore kinematically 
similar to the signal. This signal-like component remains after the forward 
jet tagging cuts.

As is readily seen from the first line of Table~\ref{WW_data}, the most 
worrisome background is $W$ pairs from $t\bar{t} + jets$ production. Of the 
1080 fb at the basic cuts level, 12 fb are from $t\bar{t}$, 310 fb are from 
$t\bar{t}j$, and the remaining 760 fb arise from $t\bar{t}jj$ production. 
The additional jets (corresponding to massless partons) are required to be 
identified as far forward tagging jets. The $t\bar{t}jj$ cross section is 
largest because the $t\bar{t}$ pair is not required to have as large an 
invariant mass as in the first two cases, where one or both $b$'s from the 
decay of the top 
quarks are required to be the tagging jets. 

For the events where one or both of the $b$'s are not identified as the 
tagging jets, they will most frequently lie between the two tagging jets, in 
the region where we search for the $W$ decay leptons. Vetoing events with 
these additional $b$ jets provides a powerful suppression tool to control the 
top background. Note that this does {\it not} require a $b$-tag, merely 
rejection of any events that have an additional jet, which in this case would 
be from a hadronically decaying $b$. We discard all events where a $b$ or 
$\bar b$ jet with $p_T > 20$~GeV is observed in the gap region between the 
tagging jets,
\bq
\label{eq:bveto}
p_{T_b} > 20 {\rm GeV} \, , \qquad
\eta_{j,min} < \eta_{b} < \eta_{j,max} \, .
\eq
This leads to a reduction of $t\bar{t}j$ events by a factor 7 while $t\bar{t}jj$ 
events are suppressed by a factor 100. This results in cross sections of 43 and 
7.6 fb, respectively, at the level of the forward tagging cuts of 
Eqs.~(\ref{eq:basic}-\ref{eq:gap}), which are now comparable to the other 
individual backgrounds. This is shown in the second line of Table~\ref{WW_data}. 
Note that the much higher $b$ veto probability for $t\bar{t}jj$ events results 
in a lower cross section than that for $t\bar{t}j$ events, an ordering which 
will remain even after final cuts have been imposed (see below).

\begin{figure}[t]
\vspace*{0.5in}
\begin{picture}(0,0)(0,0)
\includegraphics{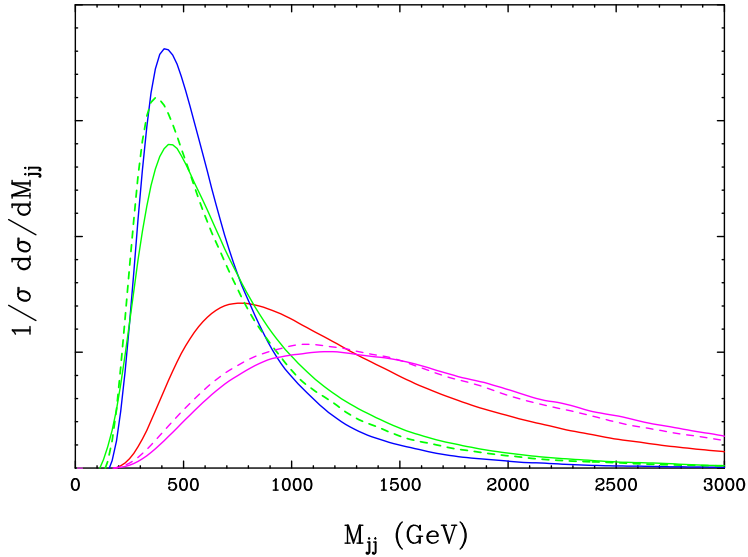}
\end{picture}
\vspace{6.5cm}
\caption{Normalized invariant mass distribution of the two tagging jets for 
the signal (red) and various backgrounds: $t\bar{t} + jets$ (blue), 
QCD $WWjj$ (solid green), EW $WWjj$ (solid purple), 
QCD $\tau\tau +jj$ (dashed green) and EW $\tau\tau +jj$ (dashed purple). 
The cuts of Eqs.~(\ref{eq:basic}-\ref{eq:bveto}) are imposed.}
\vspace*{0.2in}
\label{fig:Mjj}
\end{figure}

QCD processes at hadron colliders typically occur at smaller invariant masses 
than EW processes, due to the dominance of gluons at small Feynman $x$ in the 
incoming protons. We observe this behavior here, as shown in Fig.~\ref{fig:Mjj}. 
The three $t\bar{t} + jets$ backgrounds have been combined for clarity, even 
though their individual distributions are slightly different. We can thus 
significantly reduce much of the QCD background by imposing a lower bound on the 
invariant mass of the tagging jets: 
\bq
\label{eq:mjj}
m_{jj} > 650~{\rm GeV} \; .
\eq

\begin{figure}[t]
\vspace*{0.5in}
\begin{picture}(0,0)(0,0)
\includegraphics{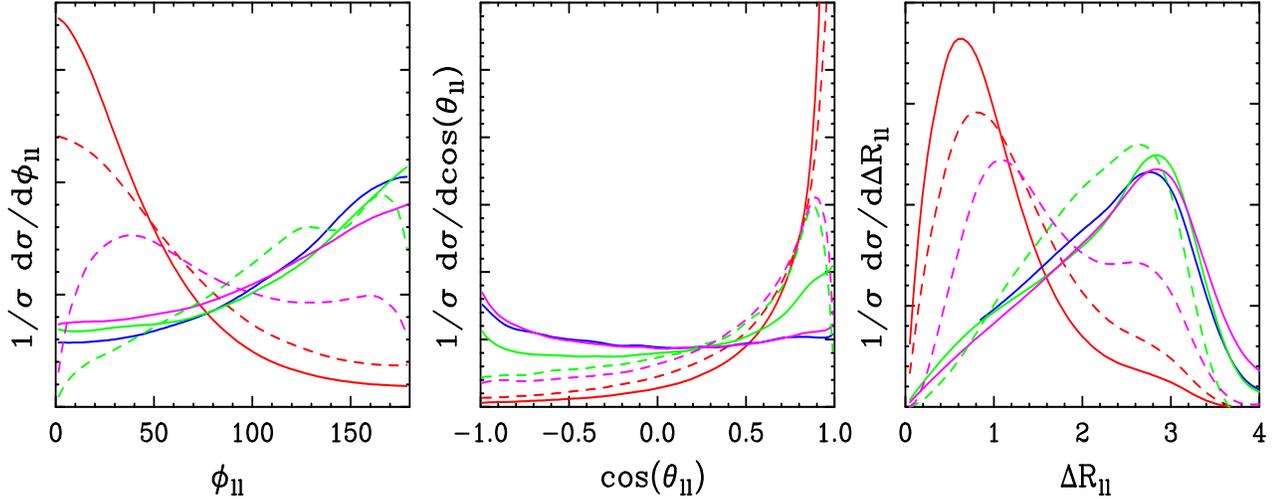}
\end{picture}
\vspace{5.5cm}
\caption{Normalized angular distributions of the charged leptons: azimuthal 
opening angle, lab opening angle, and separation in the lego plot. Results 
are shown for a Higgs boson mass of 160~GeV and 190~GeV 
(solid and dashed red lines) 
and for the various backgrounds as in Fig.~\ref{fig:Mjj}. Lepton angular 
separation is clearly smaller for the $m_H = 160$~GeV scenario. The cuts of 
Eqs.~(\ref{eq:basic}-\ref{eq:bveto}) are imposed.}
\vspace*{0.1in}
\label{fig:angdist}
\end{figure}

Another significant difference is the angular distribution of the charged 
decay leptons, $e^\pm$ and $\mu^\mp$, relative to each other. In the case 
of the Higgs 
signal, the $W$ spins are anti-correlated, so the leptons are preferentially 
emitted in the same direction, close to each other. A significant fraction of 
the various backgrounds does not have anti-correlated $W$ spins. 
These differences are demonstrated in Fig.~\ref{fig:angdist}, which shows the 
azimuthal (transverse plane) opening angle, polar (lab) opening angle, and 
separation in the lego plot. We exploit these features by establishing the 
following lepton-pair angular cuts:
\bq
\label{eq:ang}
\phi_{e\mu} < 105^{\circ} \, , \, \, \, 
{\rm cos} \; \theta_{e\mu} > 0.2 \, , \, \, \, 
\triangle R_{e\mu} < 2.2 \, .
\eq
It should be noted that while these cuts appear to be very conservative, for 
higher Higgs boson masses the $\phi_{e\mu}$ and $\triangle{R}_{e\mu}$ 
distribution broadens out to higher values, overlapping the backgrounds more. 
For $m_H \sim 180-200$~GeV these cuts are roughly optimized and further 
tightening would require greater integrated luminosity for discovery at this 
upper end of the mass range. Because of the excellent signal-to-background 
ratio achieved below, we prefer to work with uniform acceptance cuts, instead 
of optimizing the cuts for specific Higgs boson mass regions. 

\begin{figure}[t]
\vspace*{0.5in}
\begin{picture}(0,0)(0,0)
\includegraphics{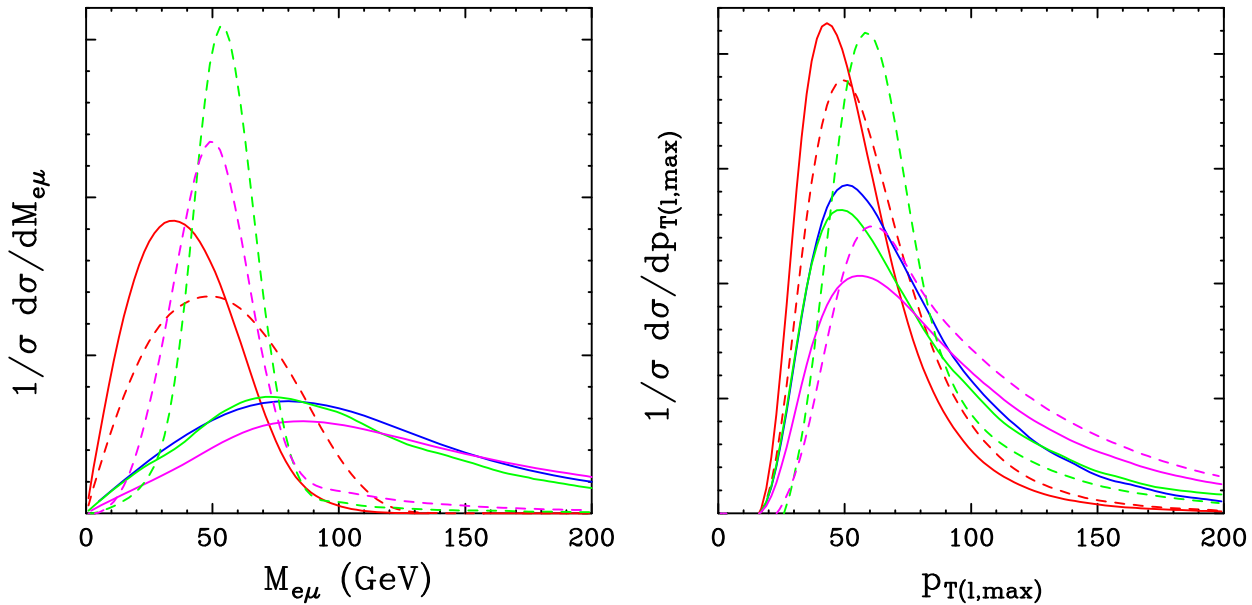}
\end{picture}
\vspace{6.5cm}
\caption{Normalized distributions of the dilepton invariant mass and maximum 
charged lepton momentum after the cuts of 
Eqs.~(\ref{eq:basic}-\ref{eq:bveto}). Results are shown for a Higgs boson 
mass of 160~GeV and 190~GeV (solid and dashed red 
lines) and for the various backgrounds as in Fig.~\ref{fig:Mjj}. The 
$m_H = 160$~GeV curve peaks at lower values of $m_{e\mu}$ and 
$p_{T_{\ell,max}}$. }
\vspace*{0.1in}
\label{fig:mllptl}
\end{figure}

We also examine the distributions for lepton-pair invariant mass, $m_{e\mu}$, 
and maximum lepton $p_T$, as shown in Fig.~\ref{fig:mllptl} for the case 
$m_H = 160,\; 190$~GeV. As is readily seen, the QCD backgrounds and EW $WWjj$ 
background prefer significantly higher values for both observables. Thus, in 
addition to the angular variables, we find it useful to restrict the 
individual $p_T$ of the leptons, as well as the invariant mass of the pair: 
\bq
\label{eq:adv}
m_{e\mu} < 110 \; {\rm GeV} \, , \, \, \, 
p_{T_{e,\mu}} < 120 \; {\rm GeV}  \, .
\eq
These are particularly effective against the top backgrounds, where the large 
top mass allows for very high-$p_T$ leptons far from the tagging jets, and 
against the EW $WWjj$ background, where the leptons tend to be well-separated 
in the lego plot. Again, the cuts are set quite conservatively so as not to 
bias a lower Higgs boson mass. Results after cuts~(\ref{eq:mjj}-\ref{eq:adv}) 
are shown on the third line of Table~\ref{WW_data}, for the case of a 160~GeV 
Higgs boson. 

\begin{table}
\caption{Signal rates $\sigma\cdot B(H\to e^\pm\mu^\mp\sla{p_T})$ for 
$m_H = 160$~GeV and corresponding background cross sections, in $pp$ collisions 
at $\protect\sqrt{s}=14$~TeV. Results are given for various levels of cuts and 
are labeled by equation numbers discussed in the text. On lines six the minijet 
veto is included. Line five gives the survival probabilities for each process, 
with $p_T^{veto} = 20$~GeV. The expected tagging jet identification efficiency 
is shown on the last line. All rates are given in fb.}
\vspace{0.15in}
\label{WW_data}
\begin{tabular}{l|cccccc|c}
cuts & $Hjj$ & $t\bar{t} + jets$ & QCD $WWjj$ & EW $WWjj$ 
     & QCD $\tau\tau jj$ & EW $\tau\tau jj$ & S/B \\
\hline
forward tagging (\ref{eq:basic}-\ref{eq:gap})
& 17.1 & 1080 & 4.4  &  3.0 & 15.8 & 0.8  & $\approx$1/65 \\
+ $b$ veto (\ref{eq:bveto})
&      &   64 &      &      &      &      & 1/5.1 \\
+ $M_{jj}$, angular cuts (\ref{eq:mjj}-\ref{eq:adv})
& 11.8 & 5.5  & 0.54 & 0.50 & 3.6  & 0.4  & 1.1/1 \\
+ real $\tau$ rejection (\ref{eq:tau})
& 11.4 & 5.1  & 0.50 & 0.45 & 0.6 &  0.08 & 1.7/1 \\
$P_{surv,20}$
& ${\it\times 0.89}$ & ${\it\times 0.29}$ & ${\it\times 0.29}$ 
& ${\it\times 0.75}$ & ${\it\times 0.29}$ & ${\it\times 0.75}$ & - \\
+ minijet veto (\ref{eq:mjveto})
& 10.1 & 1.48 & 0.15 & 0.34 & 0.18 & 0.07 & 4.6/1 \\
+ tag ID efficiency (${\it\times 0.74}$)
&  7.5 & 1.09 & 0.11 & 0.25 & 0.13 & 0.05 & 4.6/1 \\
\end{tabular}
\end{table}

At this level of cuts we now observe the combined QCD and EW $\tau\tau jj$ 
backgrounds to remain large, contributing almost $40\%$ of the total. 
We can take advantage of the fact that in these backgrounds, the $Z$ or 
$\gamma$ is emitted with quite high $p_T$, on the order 
of 100~GeV, which contributes to large $\tau$ boosts and causes the $\tau$ 
decay products to be nearly collinear in the lab frame. Within the collinear 
approximation, the $\tau$ momenta can be reconstructed knowing the charged 
lepton momenta and the missing transverse momentum 
vector~\cite{RZ_tautau,ATLAS_tau}. Labeling by $x_{\tau_1},x_{\tau_2}$ the 
fraction of $\tau$ energy each charged lepton takes with it in the $\tau$ 
decay, 
$\sla{p}_{T,x},\sla{p}_{T,y}$ can be used to solve the two equations (x,y 
transverse directions) for the two unknowns $x_{\tau_{1,2}}$. For real $\tau$ 
decays, the $\sla{p_T}$ vector must lie between the two leptons, and apart 
from finite detector resolution the reconstruction must yield 
$0 < x_{\tau_{1,2}} < 1$. For the $Hjj$ signal and other backgrounds, the 
collinear approximation is not valid because the $W$'s receive modest boosts 
in the lab only. In this case, the $\sla{p_T}$ vector will rarely lie 
between the two leptons, and an attempt to reconstruct a $\tau$ pair will 
result in $x_{\tau_1} < 0$ or $x_{\tau_2} < 0$ for $95\%$ of the 
events.~\footnote{Conversely, requiring $x_{\tau_1} > 0$, $x_{\tau_2} > 0$ 
largely eliminates $WW$ backgrounds and promises clean isolation of 
$H\to\tau\tau\to e^\pm \mu^\mp \sla{p}_T$~\cite{RPZ_prep}.} 
Additionally, the ``$\tau$ pair'' invariant mass that is reconstructed does 
not peak at $m_Z$, even when it is positive. 
We can therefore apply a highly efficient cut against the QCD and EW 
$\tau\tau jj$ backgrounds by vetoing events where an attempt to reconstruct a 
$\tau$ pair in the collinear decay approximation results in two ``real'' 
$\tau$'s near the Z pole:
\bq
\label{eq:tau}
x_{\tau_1} , \; x_{\tau_2} > 0 \; , \qquad
m_Z - 25\ {\rm GeV}\; < m_{\tau\tau} < \; m_Z + 25\ {\rm GeV} \, .
\eq
The results of this final cut are shown in line four of Table~\ref{WW_data}. 
The $\tau$ backgrounds are virtually eliminated, while the signal and the 
other backgrounds each lose $\approx 5\%$.

%%%%%%%%%%%%%%%%%%%%%%%%%%%%%%%%%%%%%%%%%%%%%%%%%%%%%%%%%%%%%%%%%%%%%%%%%%%%%%%%
%%%%%%%%%%%%%%%%%%%%%%%%%%%%%%%%%%%%%%%%%%%%%%%%%%%%%%%%%%%%%%%%%%%%%%%%%%%%%%%%

\section{Radiation patterns of minijets}
\label{sec:minijet}

If we are to veto central $b$ jets to reduce the $t\bar t + jets$ background 
to a manageable level, we must take care to correctly estimate higher-order 
additional central partonic emission in the signal and backgrounds. 
Fortunately, due to the absence of color exchange between the two scattering 
quarks in EW processes, which includes our $Hjj$ signal, we expect soft gluon 
emission mainly in the very forward and very backward directions. However, for 
QCD processes, which are dominated by $t$-channel color octet exchange, soft 
gluon radiation occurs mainly in the central detector. Thus, when we estimate 
additional central radiation with $p_T \geq 20$~GeV to match our $b$ veto 
condition, we will reject QCD background events with much higher probability 
than the EW processes. Our $b$ veto is then automatically also a minijet veto, 
a tool for QCD background suppression which has been previously studied in 
great detail for $Hjj$ production at hadron 
colliders~\cite{bpz_minijet,RZ_tautau,DZ_IZ_minijet}. 

Largely following the analysis of Ref.~\cite{RSZ_vnj} for 
the analogous EW $Zjj$ 
process which would be used to ``calibrate'' the tool at the LHC, 
we veto additional central jets in the region
\begin{mathletters}
\label{eq:mjveto}
\begin{eqnarray}
p_{Tj}^{\rm veto} \; & > & \; p_{T,{\rm veto}}\;, \label{eq:ptveto} \\
\eta_{j,min}^{\rm tag} \; & < & \eta_j^{\rm veto}
< \; \eta_{j,max}^{\rm tag} \; , \label{eq:etaveto}
\end{eqnarray}
\end{mathletters}
where $p_{T,\rm veto}$ may be chosen based on the capability of the detector. 
For the LHC we take this to be 20~GeV.

For $p_{T,\rm veto} \approx 40$~GeV we are already leaving the validity range 
of fixed-order perturbation theory in QCD processes: the cross section with 
one additional parton starts to exceed the hard tree-level cross section. As 
a result it becomes difficult to provide reliable theoretical estimates of 
minijet emission rates for the QCD backgrounds. However, gluon emission is 
governed by very different scales in signal as compared to background 
processes, due to their different color structures. Thus, a parton shower 
approach does not immediately give reliable answers unless both color 
coherence and the choice of scale are implemented correctly, matching the 
answer given by QCD matrix elements for sufficiently hard partons.

While the necessary information on angular distributions and hardness of 
additional radiation is available in the ``3-jet'' and $t\bar t + jets$ 
processes discussed in Section~\ref{sec:calc}, we must either regulate or 
reinterpret these divergent cross sections. We use the truncated shower 
approximation (TSA)~\cite{TSA} for the former, treating the ``2-jet'' cross 
sections as the inclusive rate. Details of this procedure can be found in 
Refs.~\cite{RZ_tautau,RSZ_vnj}, but here we improve upon the determination 
of veto probabilities. In our previous studies, TSA matching 
was performed without enforcement of 
the forward tagging cuts of Eqs.~(\ref{eq:lepcen},\ref{eq:gap}), even though 
tagging jet candidates were chosen for the purpose of identifying the veto 
candidate; tagging jet candidates were selected as the two most 
energetic~\cite{RSZ_vnj} or two highest-$p_T$~\cite{RZ_tautau} defined jets 
($p_T > 20$~GeV), in opposite detector hemispheres. Without the 
additional forward tagging cuts, in particular without the large rapidity 
separation of the two tagging jets, this favors QCD background events with 
high $p_T$ central quark jets which in turn lead to a harder gluon emission
spectrum than is present after all final cuts are imposed. A harder gluon
spectrum goes hand-in-hand, however, with an increased minijet emission 
probability.
A more realistic estimate of the minijet $p_T$ spectrum is obtained by 
applying the matching condition (or calculating $\bar{n}$) only in the 
phase space region where a comparison of signal and background will take 
place: after all acceptance cuts, determined at the two-jet level, have 
been imposed. 

Once the full level of cuts for a given search scenario are imposed, one may 
examine different tagging jet selection algorithms to optimize the veto. 
Ideally, the outgoing quarks would always be selected, so that the additional 
gluon radiation is always the veto candidate. In practice, this is impossible, 
but for the Higgs signal various algorithms can achieve ``proper'' quark tagging 
with about $75\%$ efficiency, a high success rate. Briefly, these might be the 
two highest-$p_T$ jets, or the two jets closest to the reconstructed Higgs. 
Most algorithms have very little difference from each other in the case of the 
WBF signature. Thus, we choose an algorithm that allows more suppression of the 
QCD backgrounds. The final algorithm we chose is to select the highest-$p_T$ jet 
as the first tagging jet, since it will almost always be part of the hard 
scattering, and then select the other tagging jet such that the event is more 
likely to pass the forward tagging cuts: look for jets with $p_T > 20$~GeV in 
the opposite hemisphere, such that the candidate Higgs decay products are 
between the tagging jets, satisfying Eq.~(\ref{eq:lepcen}). This performs 
somewhat superior to merely choosing the two highest-$p_T$ jets.

Also in contrast to our previous studies~\cite{RZ_tautau,RSZ_vnj}, the veto 
candidates are defined jets ($p_T > 20$~GeV) anywhere between the tagging jets,
{\em i.e.} they are searched for in a somewhat larger rapidity region than the 
$W$ decay leptons (see Eq.~(\ref{eq:lepcen}), 
which have to be at least 0.7 units of rapidity 
away from the tagging jets.  The choice of Eq.~(\ref{eq:etaveto}) allows for 
more suppression of the backgrounds than the more restrictive selection.
The resulting veto probabilities are summarized in line six of 
Table~\ref{WW_data}. We 
emphasize that while these probabilities are estimates only, they can be 
independently determined at the LHC in processes like $Zjj$ and $Wjj$ 
production~\cite{RSZ_vnj,CZ_gap}. 

For the $t\bar{t} + jets$ backgrounds, it is simpler instead to reinterpret the 
divergent higher-order cross sections. For this we assume that additional soft 
parton emission, which will be dominated by soft gluons, exponentiates like soft 
photon emission. This approximation has been shown well to describe multijet 
events at the Tevatron~\cite{CDFjets}. In this model, the probability to observe 
$n$ soft jets in the veto region is given by a Poisson distribution with
\bq
\bar n = \bar n(p_{T,\rm veto}) = {1 \over \sigma_2}\; 
\int_{p_{T,\rm veto}}^{\infty} dp_{T3}\; 
{d\sigma_3 \over dp_{T3}}\; ,
\eq
where the unregularized three-parton cross section is integrated over the veto 
region of Eq.~(\ref{eq:mjveto}) and then normalized to the 2-jet cross section 
$\sigma_2$, regarded as inclusive. We call this model the ``exponentiation 
model''. A rough estimate of the multiple emission probability is thus provided 
by
\begin{equation}
\label{Pvetoexp}
P_{exp}(p_{T,\rm veto}) = 1-P_0 = 1- e^{-\bar n(p_{T,\rm veto})} 
\end{equation}
which is the probability to veto the event. We find veto survival probabilities 
of $P_{surv} = 46\%$ for $t\bar{t}$ events and $P_{surv} = 12\%$ for 
$t\bar{t}j$ events. Both of these results disagree with our other estimates of 
$P_{surv}$ for QCD processes. This may be understandable for $t\bar{t}$ events, 
as at tree level this component does not contain any t-channel gluon exchange 
processes, which all of the other QCD backgrounds do. We also observe that the 
additional radiation in $t\bar{t}$ events typically falls outside the central 
gap. We did not explore this any further as the $t\bar{t}$ component is 
negligible. That the value of $P_{surv}$ found for $t\bar{t}j$ events is so 
much smaller than that for other QCD backgrounds may be understood because the 
large mass of the tops produced requires significant $p_T$, but as this is not 
yet fully explored we prefer to remain conservative and assume the value 
$P_{surv} = 0.29$ for all QCD backgrounds, including top quark pair production.
For a Higgs boson mass of 160~GeV we are left with a signal cross section of 
7.5~fb compared to a total background of 1.6~fb.

So far we have considered a single Higgs boson mass of 160~GeV only. Since we 
have largely avoided mass-specific cuts, we can immediately extend our results 
to a larger range of $m_H$. The expected number of signal events for 
115~GeV~$\leq m_H \leq 200$~GeV and an integrated luminosity of 5~fb$^{-1}$ 
are shown in Table~\ref{summary}. For the same luminosity, 8.1 background 
events are expected. In the second row of Table~\ref{summary} the Poisson 
probabilities for this background to fluctuate up to the signal level are 
given, in terms of the equivalent Gaussian significances which can be 
expected in the experiment on average.

\begin{table}
\caption{Number of expected events for the $Hjj$ signal, for 5~${\rm fb}^{-1}$ 
integrated luminosity and application of all efficiency factors and cuts, 
including a minijet veto, but for a range of Higgs boson masses. The total 
background is 8.1 events. As a measure of the Poisson probability of the 
background to fluctuate up to the signal level, the second line gives 
$\sigma_{Gauss}$,the number of Gaussian equivalent standard deviations.}
\vspace{0.15in}
\label{summary}
\begin{tabular}{c|cccccccccc}
$m_H$            & 115 & 120 & 130 & 140  & 150  & 160  & 170  & 180  & 190  & 200  \\
\hline
no. events       & 2.0 & 3.6 & 8.8 & 15.8 & 24.0 & 37.5 & 36.3 & 29.9 & 20.8 & 16.3 \\
$\sigma_{Gauss}$ & 0.5 & 1.0 & 2.6 &  4.4 &  6.3 &  9.0 &  8.8 &  7.5 &  5.5 &  4.5 \\
\end{tabular}
\end{table}

%%%%%%%%%%%%%%%%%%%%%%%%%%%%%%%%%%%%%%%%%%%%%%%%%%%%%%%%%%%%%%%%%%%%%%%%%%%%%%%%
%%%%%%%%%%%%%%%%%%%%%%%%%%%%%%%%%%%%%%%%%%%%%%%%%%%%%%%%%%%%%%%%%%%%%%%%%%%%%%%%

\section{Discussion}
\label{sec:disc}

While we have demonstrated a series of kinematic cuts and a minijet veto, based 
on the physics of the processes involved, that reduce the background well below 
the level of the signal, we have not performed a detailed detector simulation.
A full simulation is clearly needed eventually, but we do not expect our 
results to change dramatically since CMS and ATLAS will be highly efficient 
detectors. For a more realistic estimate of the signal significance we do take 
one major reduction factor into account in the last line of Table~\ref{WW_data}, 
the reconstruction efficiency for tagging jets. This is expected to be 0.86 at 
CMS for each tagging jet, resulting in a net efficiency of 
$\epsilon_{tag} = (0.86)^2 = 0.74$. Even after this nontrivial loss of total 
rate, it is clear that the method we propose still works beautifully.

\begin{figure}[htb]
\vspace*{0.5in}
\begin{picture}(0,0)(0,0)
\includegraphics{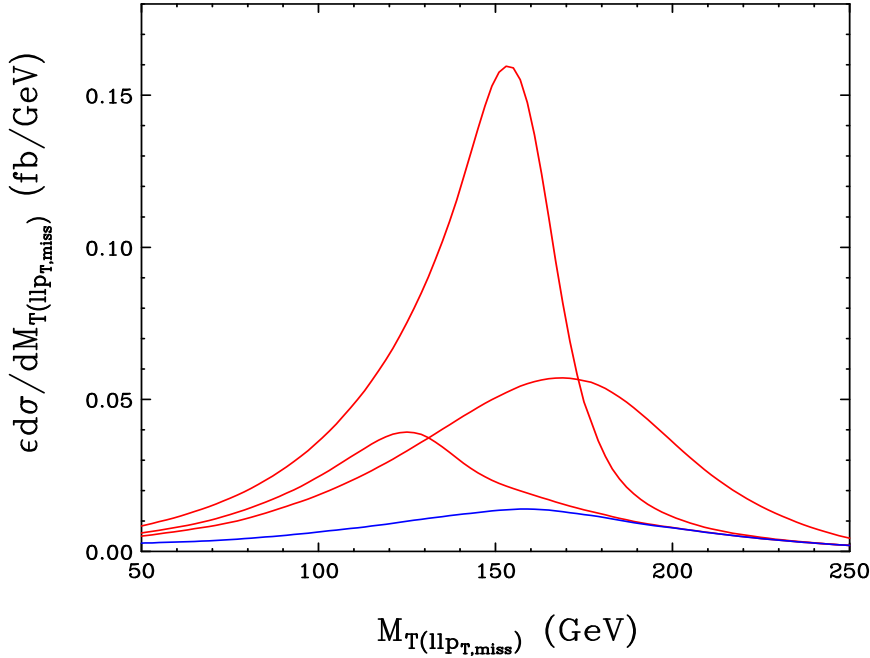}
\end{picture}
\vspace{8.5 cm}
\caption{Dilepton-$\sla{p_T}$ transverse mass distributions expected for a 
Higgs boson of mass $m_H$ = 130, 160, and 190~GeV (red) after the cuts of 
Eqs.~(\ref{eq:basic}-\ref{eq:adv}) and application of all detector efficiencies 
and a minijet veto with $p_{T,{\rm veto}} = 20$~GeV. Also shown is the 
background only (dashed).}
\label{fig:M_T}
\vspace*{0.1in}
\end{figure}

As the $H\to WW$ mode is likely to be the discovery channel for the mass range 
130~GeV~$< m_H <$~200~GeV, we wish to be able to reconstruct the Higgs boson 
mass. At threshold, the two (virtual) $W$'s are at rest in the Higgs boson 
center-of-mass frame, resulting in $m_{e\mu} = m_{\nu\bar\nu}$, so we can 
calculate the transverse energy of both the charged lepton and invisible 
neutrino systems,
\bq
\label{eq:E_T}
E_{T_{e\mu}} = \sqrt{\vec{p}_{T_{e\mu}}^2 + m_{e\mu}^2} \,  , \qquad
\sla{E}_T = \sqrt{\vec\sla{p}_T^2 + m_{e\mu}^2} \,  .
\eq
Using these results for the transverse energies, we may compute a transverse 
mass of the dilepton-$\vec\sla{p}_T$ system, 
\bq
\label{eq:M_T}
M_{T_{WW}} \; = \; 
\sqrt{(\sla{E_T}+E_{T_{e\mu}})^2 - ({\vec{p}}_{T_{e\mu}}+\vec\sla{p_T})^2}
\, ,
\eq
At threshold this is exactly the Higgs boson transverse mass. Below threshold, 
the relation $m_{e\mu} = m_{\nu\bar\nu}$ is still an excellent approximation, 
while above threshold it begins to lose validity as the $W$ bosons acquire a 
non-zero velocity in the Higgs boson rest frame. But even at $m_H=200$~GeV 
this ``pseudo'' transverse mass remains extremely useful for mass reconstruction. 
We show the dramatic results in Fig.~\ref{fig:M_T}, for Higgs boson masses of 
130, 160 and 190~GeV. Clearly visible is the Jacobian peak at $M_{T_{WW}} = m_H$, 
in particular for $m_H = 160$~GeV. The combined backgrounds are added to the 
Higgs signal, and are shown after application of all cuts and detector 
efficiencies, as well as both the $b$ and minijet vetoes which were 
discussed in the previous Sections. 

As the $M_T$ distribution shows a pronounced Jacobian peak for all Higgs 
boson masses under consideration, the signal-to-background ratio and the 
statistical significance of the signal can be improved by restricting attention
to the relevant $M_T$ range. We find that up to $W$-pair threshold 
($M_H \leq 165$~GeV), the requirement $M_T < M_H + 20$~GeV works well. 
Above threshold, loosening this cut improves the signal significance. 
Table~\ref{MTsum} 
demonstrates the power of this constraint, showing that it is possible to 
isolate a virtually background free $qq\to qqH,\;H\to WW$ signal at the LHC, 
with sufficiently large counting rate to obtain a better than $5\sigma$ signal 
with a mere 5~fb$^{-1}$ of data for the mass range 140-200~GeV. Extending the 
observability region down to 130~GeV requires at most 15~fb$^{-1}$. To reach 
120~GeV would require $\approx 65$~fb$^{-1}$ of low luminosity data 
($10^{33}\; {\rm cm}^{-2}\; {\rm s}^{-1}$), and to reach 115~GeV would require 
$\approx 165$~fb$^{-1}$. This nicely overlaps the regions of observability for 
$H\to\gamma\gamma$ (100-150~GeV)~\cite{RZ_gamgam} and $H\to\tau\tau$ 
(110-140~GeV)\cite{RZ_tautau}. The luminosity requirements in the low mass 
region can be reduced significantly by more stringent charged lepton cuts.

\begin{table}[b]
\caption{Number of expected events for the $Hjj$ signal, for 5~${\rm fb}^{-1}$ 
integrated luminosity and a range of Higgs masses. As compared to 
Table~\protect\ref{summary} an additional upper limit on the Higgs transverse 
mass of Eq.~(\ref{eq:M_T}) is imposed, as given in the first line. 
The number of both signal and background 
events are shown, as well as S/B. The Poisson probability of the background 
to fluctuate up to the signal level is given in terms of $\sigma_{Gauss}$, the 
number of Gaussian equivalent standard deviations.}
\vspace{0.15in}
\label{MTsum}
\begin{tabular}{c|cccccccccc}
$m_H$ (GeV)
& 115 & 120 & 130 & 140  & 150  & 160  & 170  & 180  & 190  & 200  \\
\hline
$M_T$ cutoff (GeV)
& 135 & 140 & 150 & 160  & 170  & 180  & 210  & 220  & none & none \\
no. S events      
& 1.9 & 3.4 & 8.3 & 14.8 & 22.7 & 36.5 & 35.9 & 29.3 & 20.8 & 16.3 \\
no. B events      
& 3.0 & 3.4 & 4.0 &  4.7 &  5.4 &  6.0 &  7.2 &  7.5 &  8.1 &  8.1 \\
S/B	 	  
& 0.6 & 1.0 & 2.0 &  3.1 &  4.2 &  6.1 &  5.0 &  3.9 &  2.6 &  2.0 \\
$\sigma_{Gauss}$  
& 0.8 & 1.4 & 3.1 &  5.0 &  6.8 &  9.6 &  9.0 &  7.6 &  5.5 &  4.5 \\
\end{tabular}
\end{table}

The high purity of the signal is made possible because 
the weak boson fusion process, together with the 
$H\to W^+ W^-\to e^\pm \mu^\mp \sla{p_T}$ decay, provides a complex signal with 
a multitude of characteristics which distinguish it from the various backgrounds. 
The basic feature of the $qq\to qqH$ signal is the presence of two forward 
tagging jets inside the acceptance region of the LHC detectors, of sizable 
$p_T$, and of dijet invariant mass in the TeV range. Typical QCD backgrounds, 
with isolated charged leptons and two hard jets, are much softer. 
In addition, the QCD backgrounds are dominated by $W$ bremsstrahlung off 
forward scattered quarks, which give typically higher-rapidity charged 
leptons. In contrast, the EW processes give rise to quite central leptons, 
and this includes not only the Higgs signal but also EW $WWjj$ and 
$\tau\tau jj$ production, which also proceed via weak boson fusion. It is 
this similarity that prevents one from ignoring EW analogs to background 
QCD processes, which a priori are smaller by two orders of magnitude in 
total cross section, but after basic cuts remain the same size as their 
QCD counterparts.

For $H\to WW$ decays, lepton angular distributions are extremely useful for 
reducing the backgrounds even further. The anti-correlation of $W$ spins in 
$H$ decay forces the charged leptons to be preferentially emitted in the same 
direction, close together in the lego plot. This happens for a small fraction 
of the background only. We have identified the most important distributions 
for enhancing the signal relative to the background, and set the various cuts 
conservatively to avoid bias for a certain Higgs boson 
mass range. There is ample room 
for improvement of our results via a multivariate analysis of a complete set 
of signal and background distributions, which we encourage the LHC 
collaborations to pursue. Additional suppression of the $t\bar{t} + jets$ 
background may be possible with $b$ identification and veto in the 
$p_{T_b} < 20$~GeV region. 

In addition to various invariant mass and angular cuts, we can differentiate 
between the $W$'s of the signal and $W,t$ backgrounds and the real $\tau$'s 
in the QCD and EW $\tau\tau jj$ backgrounds. This is possible because the 
high energy of the produced $\tau$'s makes their decay products almost 
collinear. Combined with the substantial $p_T$ of the $\tau^+\tau^-$ system 
this allows for $\tau$-pair mass reconstruction. The $W$ decays do not exhibit 
this collinearity due to their large mass, thus the angular correlation 
between the $\sla{p_T}$ vector and the charged lepton momenta is markedly 
different. Our real-$\tau$ rejection makes use of these differences and 
promises to virtually eliminate the $\tau\tau jj$ backgrounds.

We advocate taking advantage of an additional fundamental characteristic of 
QCD and EW processes. Color-singlet exchange in the $t$-channel, as 
encountered in Higgs boson production by weak boson fusion (and in the EW 
$Zjj$ background), leads to additional soft jet activity which differs 
strikingly from that expected for the QCD backgrounds in both geometry and 
hardness: gluon radiation in QCD processes is typically both more central 
and harder than in WBF processes. We exploit this radiation, via a veto on 
events with central minijets of $p_T > 20$~GeV, and expect a typical $70\%$ 
reduction in QCD backgrounds and about a $25\%$ suppression of EW backgrounds, 
but only about a $10\%$ loss of the signal.

Beyond the possibility of discovering the Higgs boson in the $H\to WW$ mode, 
or confirmation of its existence, measuring the cross sections in both weak 
boson and gluon fusion will be important both as a test of the Standard Model 
and as a search for new physics. For such a measurement, via the analysis 
outlined in this paper, minijet veto probabilities must be precisely known. 
For calibration purposes, one can analyze $Zjj$ events at the LHC. The 
production rates of the QCD and EW $Zjj$ events can be reliably predicted 
and, thus, the observation of the $Z\to\ell\ell$ peak allows for a direct 
experimental assessment of the minijet veto efficiencies, in a kinematic 
configuration very similar to the Higgs signal.

Observation of SM $H\to e^\pm\mu^\mp\sla{p_T}$ at the LHC is possible for 
very low integrated luminosities, if the Higgs boson lies in the mass range 
between about 130 and 200 GeV. Weak boson fusion at the LHC will be an 
exciting process to study, for a weakly coupled Higgs sector just 
as much as for strong interactions in the symmetry breaking sector of 
electroweak interactions.

%%%%%%%%%%%%%%%%%%%%%%%%%%%%%%%%%%%%%%%%%%%%%%%%%%%%%%%%%%%%%%%%%%%%%%%%%%%%%%%%
%%%%%%%%%%%%%%%%%%%%%%%%%%%%%%%%%%%%%%%%%%%%%%%%%%%%%%%%%%%%%%%%%%%%%%%%%%%%%%%%

\acknowledgements
This research was supported in part by the University of Wisconsin Research
Committee with funds granted by the Wisconsin Alumni Research Foundation and
in part by the U.~S.~Department of Energy under Contract
No.~DE-FG02-95ER40896.

%%%%%%%%%%%%%%%%%%%%%%%%%%%%%%%%%%%%%%%%%%%%%%%%%%%%%%%%%%%%%%%%%%%%%%%%%%%%%%%%
%%%%%%%%%%%%%%%%%%%%%%%%%%%%%%%%%%%%%%%%%%%%%%%%%%%%%%%%%%%%%%%%%%%%%%%%%%%%%%%%


\newpage
\begin{references}

\bibitem{EWfits}
For recent reviews, see e.g. 
J.L.~Rosner, Comments Nucl. Part. Phys. {\bf 22}, 205 (1998) [hep-ph/9704331];
%%CITATION = CNPPA,22,205;%%
K.Hagiwara, Ann. Rev. Nucl. Part. Sci. 1998, 463; 
W.J.~Marciano, hep-ph/9902332; and references therein.
%%CITATION = HEP-PH 9902332;%%

\bibitem{reviews}
For recent reviews, see e.g. S.~Dawson, [hep-ph/9703387]; 
%%CITATION = HEP-PH 9703387;%%
M.~Spira, Fortsch.\ Phys.\ {\bf 46}, 203 (1998); 
%%CITATION = FPYKA,46,203;%%
and references therein.

\bibitem{ttH}
W.~J.~Marciano and F.~E.~Paige, 
Phys.\ Rev.\ Lett.\ {\bf 66}, 2433 (1991);
%%CITATION = PRLTA,66,2433;%%
J.~F.~Gunion, Phys.\ Lett.\ {\bf B261}, 510 (1991).
%%CITATION = PHLTA,B261,510;%%

\bibitem{WH}
A.~Stange, W.~Marciano, and S.~Willenbrock, 
Phys.\ Rev.\ {\bf D50}, 4491 (1994), [hep-ph/9404247];
%%CITATION = PHRVA,D50,4491;%%
R.~Kleiss, Z.~Kunszt, W.~J.~Stirling, 
Phys.\ Lett.\ {\bf B253}, 269 (1991);
%%CITATION = PHLTA,B253,269;%%
H.~Baer, B.~Bailey, J.~F.~Owens, 
Phys.\ Rev.\ {\bf D47}, 2730 (1993).
%%CITATION = PHRVA,D47,2730;%%

\bibitem{glo_will_hww}
E.W.~Glover, J.~Ohnemus and S.S.~Willenbrock, 
Phys.\ Rev.\ {\bf D37}, 3193 (1988);
%%CITATION = PHRVA,D37,3193;%%
V.~Barger, G.~Bhattacharya, T.~Han and B.A.~Kniehl,
Phys.\ Rev.\ {\bf D43}, 779 (1991).
%%CITATION = PHRVA,D43,779;%%

\bibitem{DittDrein}
M.~Dittmar and H.~Dreiner, 
Phys.\ Rev.\ {\bf D55}, 167 (1997); 
%%CITATION = PHRVA,D55,167;%%
and [hep-ph/9703401].
%%CITATION = HEP-PH 9703401;%%

\bibitem{bpz_minijet}
V.~Barger, R.~J.~N.~Phillips, and D.~Zeppenfeld,
Phys.\ Lett.\ {\bf B346}, 106 (1995).
%%CITATION = PHLTA,B346,106;%%

\bibitem{Cahn}
R.~N.~Cahn, S.D.~Ellis, R.~Kleiss and W.J.~Stirling, 
Phys.\ Rev.\ {\bf D35}, 1626 (1987);
%%CITATION = PHRVA,D35,1626;%%
V.~Barger, T.~Han, and R.~J.~N.~Phillips, Phys.\ Rev.\ {\bf D37}, 2005 (1988);
%%CITATION = PHRVA,D37,2005;%%
R.~Kleiss and W.~J.~Stirling, Phys.\ Lett.\ {\bf 200B}, 193 (1988);
%%CITATION = PHLTA,200B,193;%%
D.~Froideveaux, in {\it Proceedings of the ECFA Large Hadron
Collider Workshop}, Aachen, Germany, 1990, edited by G.~Jarlskog and D.~Rein
(CERN report 90-10, Geneva, Switzerland, 1990), Vol~II, p.~444;
M.~H.~Seymour, {\it ibid}, p.~557;
U.~Baur and E.~W.~N.~Glover, Nucl.\ Phys.\ {\bf B347}, 12 (1990);
%%CITATION = NUPHA,B347,12;%%
%U.~Baur and E.~W.~N.~Glover,
Phys.\ Lett.\ {\bf B252}, 683 (1990).
%%CITATION = PHLTA,B252,683;%%

\bibitem{BCHP}
V.~Barger, K.~Cheung, T.~Han, and R.~J.~N.~Phillips,
Phys.\ Rev.\ {\bf D42}, 3052 (1990);
%%CITATION = PHRVA,D42,3052;%%
V.~Barger {\it et al.},
%, K.~Cheung, T.~Han, J.~Ohnemus, and D.~Zeppenfeld,
Phys.\ Rev.\ {\bf D44}, 1426 (1991);
%%CITATION = PHRVA,D44,1426;%%
V.~Barger, % {\it et al.},
K.~Cheung, T.~Han, and D.~Zeppenfeld,
Phys.\ Rev.\ {\bf D44}, 2701 (1991);
%%CITATION = PHRVA,D44,2701;%%
erratum Phys.\ Rev.\ {\bf D48}, 5444 (1993);
%%CITATION = PHRVA,D48,5444;%%
Phys.\ Rev.\ {\bf D48}, 5433 (1993);
%%CITATION = PHRVA,D48,5433;%%
V.~Barger {\it et al.},
%, K.~Cheung, T.~Han,  A.~Stange, and D.~Zeppenfeld,
Phys.\ Rev.\ {\bf D46}, 2028 (1992).
%%CITATION = PHRVA,D46,2028;%%

\bibitem{DGOV}
D.~Dicus, J.~F.~Gunion, and R.~Vega,
Phys.\ Lett.\ {\bf B258}, 475 (1991);
%%CITATION = PHLTA,B258,475;%%
D.~Dicus, J.~F.~Gunion, L.~H.~Orr, and R.~Vega,
Nucl. \ Phys.\ {\bf B377}, 31 (1991).
%%CITATION = NUPHA,B377,31;%%

\bibitem{bjgap}
Y.~L.~Dokshitzer, V.~A.~Khoze, and S.~Troian, in {\it
Proceedings of the 6th International Conference on Physics in Collisions},
(1986) ed.\ M.~Derrick (World Scientific, 1987) p.365;
J.~D.~Bjorken, Int.\ J.\ Mod.\ Phys.\ {\bf A7}, 4189 (1992);
%%CITATION = IMPAE,A7,4189;%%
Phys.\ Rev.\ {\bf D47}, 101 (1993).
%%CITATION = PHRVA,D47,101;%%

\bibitem{TSA}
V.~Barger and R.~J.~N.~Phillips, Phys.\ Rev.\ Lett.\ {\bf 55}, 2752 (1985);
%%CITATION = PRLTA,55,2752;%%
H.~Baer, V.~Barger, H.~Goldberg, and R.~J.~N.~Phillips,
Phys.\ Rev.\ {\bf D37}, 3152 (1988).
%%CITATION = PHRVA,D37,3152;%%

\bibitem{CDFjets}
D.~Rainwater, D.~Summers, and D.~Zeppenfeld,
Phys.\ Rev.\ {\bf D55}, 5681 (1997).
%%CITATION = PHRVA,D55,5681;%%

\bibitem{CTEQ4_pdf}
H.~L.~Lai {\it et al.}, Phys.\ Rev.\ {\bf D55}, 1280 (1997), [hep-ph/9606399].
%%CITATION = PHRVA,D55,1280;%%

\bibitem{qqHorig}
R.~Cahn and S.~Dawson,  Phys.\ Lett.\ {\bf 136B}, 196 (1984).
%%CITATION = PHLTA,136B,196;%%

\bibitem{RZ_gamgam}
D.~Rainwater and D.~Zeppenfeld, Journal of High Energy Physics 12, 005 (1997).
%%CITATION = JHEPA,9712,005;%%

\bibitem{RZ_tautau}
D.~Rainwater, D.~Zeppenfeld and K.~Hagiwara, 
Phys.\ Rev.\ {\bf D59}, 014037 (1999).
%%CITATION = PHRVA,D59,014037;%%

\bibitem{RSZ_vnj}
D.~Rainwater, R.~Szalapski, and D.~Zeppenfeld,
Phys.~Rev.~{\bf D54}, 6680 (1996), [hep-ph/9605444].
%%CITATION = PHRVA,D54,6680;%%

\bibitem{Madgraph}
T.~Stelzer and W.~F.~Long, 
Comp.\ Phys.\ Comm.\  {\bf 81}, 357 (1994), [hep-ph/9401258].
%%CITATION = CPHCB,81,357;%%

\bibitem{Stange}
A.~Stange, private communication.

\bibitem{VVjj}
V.~Barger, T.~Han, J.~Ohnemus and D.~Zeppenfeld,
Phys.\ Rev.\ {\bf D41}, 2782 (1990).
%%CITATION = PHRVA,D41,2782;%%

\bibitem{DZ_IZ_minijet}
A.~Duff and D.~Zeppenfeld, Phys.\ Rev.\ {\bf D50}, 3204 (1994);
%%CITATION = PHRVA,D50,3204;%%
K.~Iordanidis and D.~Zeppenfeld, 
Phys.\ Rev.\ {\bf D57}, 3072 (1998), [hep-ph/9709506].
%%CITATION = PHRVA,D57,3072;%%

\bibitem{Kst}
S.~D.~Ellis, R.~Kleiss, and W.~J.~Stirling,
Phys.\ Lett.\ {\bf 154B}, 435 (1985);
%%CITATION = PHLTA,154B,435;%%
R.~Kleiss and W.~J.~Stirling, Nucl.\ Phys.\ {\bf B262}, 235 (1985);
%%CITATION = NUPHA,B262,235;%%
Phys.\ Lett.\ {\bf 180B}, 171 (1986);
%%CITATION = PHLTA,180B,171;%%
J.~F.~Gunion, Z.~Kunszt, and M.~Soldate, Phys.\ Lett.\ {\bf 163B}, 389 (1985);
%%CITATION = PHLTA,163B,389;%%
Erratum, Phys.\ Lett.\ {\bf 168B}, 427 (1986);
%%CITATION = PHLTA,168B,427;%%
J.~F.~Gunion and M.~Soldate, Phys.\ Rev.\ {\bf D34}, 826 (1986);
%%CITATION = PHRVA,D34,826;%%
R.~K.~Ellis and R.~J.~Gonsalves, in {\it Proc.\ of the Workshop on super high
energy physics}, Eugene, OR (1985), ed.\ D.~E.~Soper, p.~287.

\bibitem{HZ}
K.~Hagiwara and D.~Zeppenfeld, Nucl. Phys. {\bf B313}, 560 (1989).
%%CITATION = NUPHA,B313,560;%%

\bibitem{BHOZ}
V.~Barger, T.~Han and J.~Ohnemus, and  D.~Zeppenfeld,
Phys.~Rev.~Lett.~{\bf 62}, 1971 (1989); Phys.~Rev.~{\bf D40}, 2888 (1989).
%%CITATION = PHRVA,D40,2888;%%

\bibitem{BG}
F.~A.~Berends {\it et al.},
%W.~T.~Giele, H.~Kuijf, R.~Kleiss, and W.~J.~Stirling,
Phys.~Lett.~{\bf B224}, 237 (1989).
%%CITATION = PHLTA,B224,237;%%

\bibitem{CZ_gap}
H.~Chehime and D.~Zeppenfeld, Phys.\ Rev.\ {\bf D47}, 3898 (1993).
%%CITATION = PHRVA,D47,3898;%%

\bibitem{CMS-ATLAS}
W.~W.~Armstrong {\it et al.}, 
Atlas Technical Proposal, report CERN/LHCC/94-43 (1994).

\bibitem{ATLAS_tau}
D.~Cavalli, L.~Cozzi, L.~Perini, S.~Resconi,
ATLAS Internal Note PHYS-NO-051, Dec. 1994.

\bibitem{RPZ_prep}
D.~Rainwater, T.~Plehn, and D.~Zeppenfeld, preprint MADPH-99-1142
[hep-ph/9911385].

\end{references}
\end{document}